\documentclass[11pt]{article}
  \usepackage{amssymb}
     \usepackage{graphicx}
\newcommand{\sect}[1]{\setcounter{equation}{0}\section{#1}}
\renewcommand{\theequation}{\arabic{section}.\arabic{equation}}
\def\am{angular momentum~}

\def\al {\alpha}

\def\ba{\begin{eqnarray}}

\def\be{\begin{equation}}
\def\bi{\bibitem}
\def\bm{\boldmath}

\def\cd{\cdot}

\def\coo{coordinates~}
\def\cy{cylindrical~}

\def\de{\delta}

\def\di{\partial}

\def\ea{\end{eqnarray}} 
\def\ee{\end{equation}}
\def\eee{equation~}
\def\eeee{equations~}

\def\ep{\epsilon} 
\def\eq{\equiv~}

\def\ei{Einstein \eee~}
\def\eii{Einstein \eeee~}

\def\fr{\frac}

 \def\ga{\gamma}
 \def\gmn{g_{\mu\nu}}

\def\gw{gravitational wave~}
\def\gww{gravitational waves~}

\def\Ga{\Gamma}
\def\GR{General Relativity~}

\def\ha{\frac{1}{2}~}
\def\hy{hypersurface~}
\def\hyo{hypersurface orthogonal~}

\def\iii{inertial frame~}
\def\iiii{inertial frames~}
\def\inf{\infty}

\def\ka{\kappa}

\def\kv{Killing vector~}
\def\la {\lambda}
\def\lb{\label} 

\def\lhs{left-hand side~}
\def\lll{\left(}

\def\LLL{\left[}

\def\mb{\mbox}

\def\mn{\mu\nu}

\def\na{\nabla}

\def\nnn{\noindent}
\def\np{\newpage}

\def \om{ \omega}

\def\ppp{perturbation~}

\def\ps{\psi}
\def\ra{\rightarrow}
\def\rh{\rho}
\def\rhs{right-hand side~}
\def\ro{rotation~}
\def\rog{rotating~}
\def\rrr{\right)}

\def\RR{{\cal R}}
\def\Ra{\Rightarrow}
\def\RRR {\right]}

\def\si{\sigma}
\def\sq{\sqrt}
\def\sqg{\sqrt{-g}}

\def\sim{\simeq}
\def\sss{spacetime~}

\def\sy{symmetric~}

\def\td{\tilde}

\def\tr{\tilde\rho}

\def\ts{\textstyle}
\def\tt{\tilde t}

\def\TT{{\cal T}}

\def\ubm{\unboldmath}

\def\vf{\varphi }

\def\vna{\vec{\na}}
\def\vs{\vskip 0.5 cm}

\def\wrt{with respect to~}

\def\ze{\zeta}

\def\1k{\fr{1}{\ka}}

\def\2k{\fr{1}{2\ka}}

\title{{\bf Inertial frame rotation induced by rotating \gww}}
\author{  Donald Lynden-Bell$^{1}$\thanks{email:
dlb@ast.cam.ac.uk} \, Ji\v r\'\i  \,\,Bi\v c\'ak$^{2,1}$\thanks{email:bicak@mbox.troja.mff.cuni.cz
} \,    and Joseph Katz$^{3,1} $\thanks{email:jkatz@phys.huji.ac.il
  } \\
\\{\it $^1$Institute of Astronomy, Madingley Road, Cambridge CB3 0HA,
United Kingdom}\\
\\ {\it $^2$Institute of Theoretical Physics,  Charles University, 180 00 Prague 8, Czech Republic} \\
\\
 $^3${\it The Racah Institute of Physics, Givat Ram, 91904 Jerusalem,
Israel}
\\ }
 
\begin{document}
 
\maketitle

\begin{abstract} 
\setlength{\baselineskip}{20pt plus2pt}
We calculate the \ro of the \iii within an almost flat \cy region surrounded by a pulse of non-axially-\sy \gww that rotate about the axis of our \cy polar coordinates. Our \sss has only one Killing vector. It is along the $z$-axis and \hy orthogonal. We solve the \eii to first order in the wave amplitude and superpose such linearized solutions to form a wave pulse. We then solve the relevant \ei to second order in the amplitude to find the \ro of \iiii produced by the pulse. The \ro is without time delay.

The influence of \gw \am on the \iii demonstrates that Mach's principle can not be expressed in terms of the influence of the stress-energy-momentum tensor alone but must involve also influences of \gw energy and angular momentum.
\vs
\vs
\nnn  PACS numbers 04.20.Jb \,\,\, 04.30. -w

\end{abstract} 

 \np
\setlength{\baselineskip}{20pt plus2pt}
\sect{Introduction}
 Mach  \cite{Mach} rejected Newton's  \cite{Ne} absolute space and suggested that some average of the positions and motions of the masses in the Universe determines the inertial frame.
 
 Einstein \cite{Ei17}, \cite{Ei50} was inspired by Mach's principle and his earlier work \cite{Ei13} showed that the local \iii was not absolute but was affected by the rotation and acceleration of matter nearby. Extrapolating from this he said, in essence,   the relativity of inertia is not fulfilled if inertia is affected by the presence of matter it must be entirely caused by it. He also said there is no inertia of mass against space but only inertia of mass against mass. Thus he hoped that the \iii might be determined by the distribution of the stress tensor\footnote{Indices $\la,
\mu, \nu, \rho,\cdots=0,1,2,3 $; indices
$k, l, m, n\cdots=1,2,3$ and $a,b,c,\cdots=0,1,2$. The metric
$\gmn$ has signature $+---$ and $g$ is its determinant. Covariant derivatives are indicated by a $D$, partial
derivatives   by a $\di$ and covariant derivatives in a three subspace by $\na_a$ or $\na_k$. The permutation symbol in 4 dimensions is $\ep_{\mu\nu\rho\si}$ with
$\ep_{0123}=1$ and in 3 dimensions   $\ep_{klm}$ with $\ep_{123}=1$. Finally $\eta_{\mu\nu\rh\si}=\sqg \ep_{\mu\nu\rho\si} $.} $T_{\mn}$. His efforts to remove the external influence of the boundary conditions at infinity on the solutions of his \eeee led him to consider closed spaces with no spatial boundaries \cite{Ei17}. Here we show that within General Relativity, any general statement of Mach's principle that attributes all dragging of inertial frames solely to the distribution of  $T_{\mn}$ as the origin of inertia is false (see \cite{Bo52}). One counter example suffices to demonstrate this.
We find that there is an almost flat \cy region near the axis of a revolving \gw pulse (which inevitably has no $T_{\mn}$) and demonstrate that the \iii in the \cy interior rotates relative to the \iii at great distances.

The exact \cy wave solutions \cite{Be}, \cite{ER}, \cite{Bonnor}, \cite{WW} carry no \am about the axis, because their wave fronts have normals with no component around the axis. Bondi \cite{Bondi} emphasises this point.
 
 In a sense our work pushes at an open door, since now all accept that \gww exist and carry both energy and angular momentum. This is not merely known theoretically, but also observationally   \cite{TW}, \cite{TFM}, cf. also \cite{ML+}   and \cite{Ly}.

It would be astonishing if that energy and \am did not produce gravity and \ro of the inertial frame. Indeed, as we described in more detail in paper \cite{BKL07}, the work of   Corvino \cite{Co} demonstrates that this must be the case and in    \cite{BKL07}  we have explicitly calculated the gravity of a non-linear pulse of gravitational waves.

Here our aim is to produce a nice clean example of the \ro of the \iii in an almost flat region surrounded by rotating gravitational waves.

An extreme example of inertia   due to \gww alone is provided by Gowdy's \cite{Go} universe, a closed world that contains nothing except gravitational waves. One of our ultimate aims is to discuss the meaning of Mach's principle in \GR and particularly its application to such systems. However, such applications and the question as to what bearing Mach's principle has on the existence of dark energy are outside the scope of this paper.

When we first discussed Mach's principle in FRW universes via \ppp theory \cite{LKB}, we pointed out that our first order \ppp theory inevitably missed the second order contribution expected from gravitational waves (see also our recent comprehensive paper \cite{BKL}). Here we show that there must be such a contribution.

We consider waves in spaces that have a \hyo \kv along $Oz$. If such waves have axial symmetry their wave normals converge on the axis and they carry no angular momentum. To get truly \rog waves that carry \am around the axis, we can not impose axial symmetry. As no exact non-axisymmetric  \rog wave solutions are known, we work via \ppp theory. First we solve the flat space wave \eee for linearized \gww with symmetry along $Oz$, using \cy polar coordinates. Then we superpose these linear solutions to form a wave pulse  that rotates around the axis. This pulse generalizes to non-axial-symmetry the wave pulse found by Bonnor \cite{Bonnor} and by Weber and Wheeler \cite{WW}. In the \sss generated by such a pulse of linear waves, we then solve one of Einstein's \eeee to second order in the wave amplitude. Following \cite{ABS} and   \cite{BKL07}, we see that the wave pulse generates an ``effective source  term" of second order in the wave amplitude. This means that the first order changes in the metric due to the wave can be neglected when calculating   the extra \ppp due to these effective sources. (Keeping the changes could only lead to terms of third order in the wave amplitude.) Thus we have to calculate in flat space the effects of sources that are quadratic in the wave amplitude. Such perturbative techniques are well discussed in \cite{GNPP}. 
 
\sect{Calculations}
 We consider as in   \cite{BKL07} source free metrics with   one  hypersurface orthogonal Killing vector of 
  spacelike translations \bm$\ze$\ubm. Coordinates can be   chosen so that $z=x^3$ and 
  \be
  \{x^\mu\}=\{x^a, x^3\}~~~{\rm with}~~~a =0,1,2 ~~~{\rm  ~that}~~~\mb{\bm$\ze$\ubm}=\{0, 0, 0, 1\},
  \lb{21}
  \ee
and the metric can be written in the form  introduced by Ashtekar, Bi\v c\'ak and Schmidt \cite{ABS}:    
  \be
ds^2=   e^{-2\psi}g_{ab}dx^adx^b-e^{2\psi} dz^2,
\lb{22}
\ee  
$\psi$ and $g_{ab}$ are functions of the $x^c$ only. The source-free Einstein's \eeee $R_{\mn}=0$ take then the  following   form in terms of the Ricci tensor $\RR_{ab}$ of the 2+1-space $d\si^2=g_{ab}dx^adx^b$:
 \ba
 R_{ab}&=&0~~~\Ra~~~\RR_{ab}=2\di_a\psi\di_b\psi,
 \lb{23}
 \\
 R_{33}&=&0~~~\Ra~~~g^{ab}\na_a \na_b\psi=0.
 \lb{24}
 \ea 
Because $\ps$ satisfies (\ref{24}),  the \rhs of (\ref{23})
 automatically satisfies the Bianchi identities in $2+1$ spacetime. To zero order $g_{ab}$ reduces to the flat \sss metric and we choose \cy \coo $x^0=t, x^1=\rh, x^2=\vf$, so, in the notations of \cite{BKL07} where a dot means a derivative \wrt $t$  and a prime a derivative \wrt $\rh$, our linearized waves obey the wave \eee
 \be
\ddot\ps - \ps''-\fr{1}{\rh}\ps'-\fr{1}{\rh^2}\di_\vf^2\ps=0.
\lb{25}
\ee 
The solutions are expressed in Bessel functions $J_m(\om\rh)$:
\be
\ps=Ae^{\pm im\vf \pm i\om t}J_m(\om\rh).
\lb{26}
\ee
We note that the special class of these waves with $m=0$ constitute the exact solutions mentioned in the Introduction. We follow \cite{Bonnor} and \cite {WW} in choosing a superposition of such waves but in order to have   \rog linearized waves we need $m\ne0$ and we choose waves of the form $e^{i(m\vf-\om t)}$ with $m\om>0$ so that the wave fronts rotate positively about the $z$ axis. Our linear wave pulse is given by the superposition with $a>0$ being the effective duration of the pulse,
\be
\ps=B\int_0^\inf (a \om)^m e^{-a \om} e^{i(m\vf-\om t)}J_m( \om \rh)ad \om+c.c.,
\lb{27}
\ee
$B$ is a dimensionsless amplitude.
It is convenient to define $\al=\al(t)=a+it$.  Thanks to the Bateman Manuscript Project of Erd\'elyi {\it et al} \cite{EMOT}, formula 8.6.5, we find
\be
\ps=Ba^{m+1}e^{im\vf}\cd 2^m \fr{\Ga(m+\ts{\fr{1}{2}})}{\sq{\pi}}  \cd\rh^m\lll \al^2+\rh^2  \rrr^{-m-\ha}+ c.c.
\lb{28}
\ee 
   $\ps$ may be written in terms of   non-dimensional variables as in \cite{BKL07},
 \be 
 \tr=\fr{\rh}{a}~~~,~~~\tt=\fr{t}{a},~~~ {\rm and~with}~~~~2^m\fr{\Ga(m+\ts{\fr{1}{2}})}{{\sq{\pi}}}=(2m-1)!!,
 \lb{29}
 \ee
$\ps$ can also be written in real terms as follows:
\be
\ps=2B(2m-1)!!\fr{\tr^m\cos\LLL m\vf-\lll m+\ts{\fr{1}{2}}\rrr\chi\RRR}{\LLL  \lll  1+\tr^2 - \tt^2 \rrr^2+4\tt^2  \RRR^{\fr{1}{2}(m+\fr{1}{2})}}, ~~~{\rm where}~~~\chi=\arctan\fr{2\tt}{   1+\tr^2 - \tt^2}.
\lb{210}
\ee
So the phase of $\ps$ is $m\vf-\lll m+\ts{\fr{1}{2}}\rrr\chi$. The $\tr$-dependence of this phase gives spirality to the wave and changes sign at $t=0$. Figure 1 shows the profile of   $\ps$ at different times for $m=10$ and $\vf=0$.

To calculate the source term on the \rhs of (\ref{23}) we need derivatives of expression (\ref{28}). In particular the $t,\vf$ component of the source will be given by $2\dot\ps\di_\vf\ps$. This product contains two terms that are independent of $\vf$ and two terms that vary like $\sin 2m\vf$ or $\cos 2m\vf$. Those terms can not cause any \ro of the \iii on the axis. We therefore concentrate on the terms that are independent of $\vf$ and denote them by the averaging symbol $<~ >$, implying an average over $\vf$. Defining $j$, we find  from (\ref{28}) and (\ref{29}),  
\be
  j=-2<\dot\ps\di_\vf\ps>= 
\fr{2B^2}{a}m(2m+1)\LLL (2m-1)!!  \RRR^2\tr^{2m}Q^{-m-3/2}\lll \fr{\di Q}{\di u}+4\tt^2\rrr,
\lb{211} 
\ee
in which
\be
u=\tr^2~~~{\rm and}~~~ Q=\lll  1+u-\tt^2   \rrr^2 + 4\tt^2.
\lb{212}
\ee

Before proceeding further we return to \eeee (\ref{23}), now regarding the \rhs as a known second order source of \ppp of our now first order metric. As the source terms are already of second order, we may neglect the change of the metric from flat space in calculating their effects, as such changes could only give effects of third order. We can write the fully perturbed metric in the   form (\ref{22}) in which, see (5.4) in \cite{BKL07},
\be
g_{ab}dx^adx^b=e^{2\ga}\lll  dt^2-d\rh^2 \rrr -W^2(d\vf- \om dt)^2.
\lb{213}
\ee
The flat space \eee for the axially symmetrical part of $ \om$ is then, cf. (5.10) in
\cite{BKL07},
\be
<\RR_{02}>=\fr{1}{2\rh}(\rh^3<\!\om\!>')'= 2<\!\dot\ps\di_\vf\ps\!>=-j.
\lb{214}
\ee
We first integrate once with the boundary condition that $<\!\! \om\!\!>$ is not singular on the  axis:
\be
\rh^3<\!\! \om\!\!>'=  - \int_0^{\tr^2}jd\rh^2_1.
\lb{215}
\ee
Using the boundary condition that $<\!\! \om\!\!>\ra0$ at infinity we then find that
\be
<\!\! \om\!\!>= {\ts\fr{1}{2}} \int_{\tr^2}^\inf \fr{1}{\rh_2^4}\lll     \int_0^{\tr^2} jd\rh^2_1   \rrr d\rh^2_2,
\lb{216}
\ee
where $\rh^2_1$ and $\rh^2_2$ are the dummy variables. Doing the outer integration by parts, we deduce
\be
<\!\! \om\!\!>=  \fr{1}{2u}\int_0^ujdu +  \int_u^\inf\fr{1}{2u}j du.
\lb{217}
\ee
Substituting (\ref{211}) into (\ref{217}), we get, after integrating the $Q^{-m-3/2}(\di Q/\di u)$ terms by parts,
\be
<\!\! \om\!\!>= \fr{2B^2}{a} m\LLL  (2m-1)!! \RRR^2 \LLL   \fr{m}{u}I_{m-1}+2(2m+1)\fr{\tt^2}{u}I_m+(m-1)H_{m-1}+2(2m+1)\tt^2H_m  \RRR,
\lb{218}
\ee
in which    
\be
I_m(u)=\int_0^u u^mQ^{-m-3/2}du ~~~,~~~H_m(u)=\int_u^\inf u^{m-1}Q^{-m-3/2}du. 
\lb{219}
\ee
 The evaluation of $<\!\! \om\!\!>$ is thus reduced to the calculation of $I_m$ and $H_m$ for all $m$. This is done in the Appendix.

\sect{Rotation of \iiii at small and great distances}

On axis the first two terms in the large brackets of (\ref{218}) are equal to zero. So using (\ref{A14}) with $b=1-\tt^2$ and $c=1+\tt^2$ we get from (\ref{218})
\be
<\!\!\om\!\!>_0=\fr{B^2}{a} \cd\fr{(2m)!}{2^{2m-1}}\cd\fr{1+m(1+\tt^2)}{(1+\tt^2)^2}.
\lb{31}
\ee
\begin{figure}[h]
   \centering
      \includegraphics[width=15 cm]{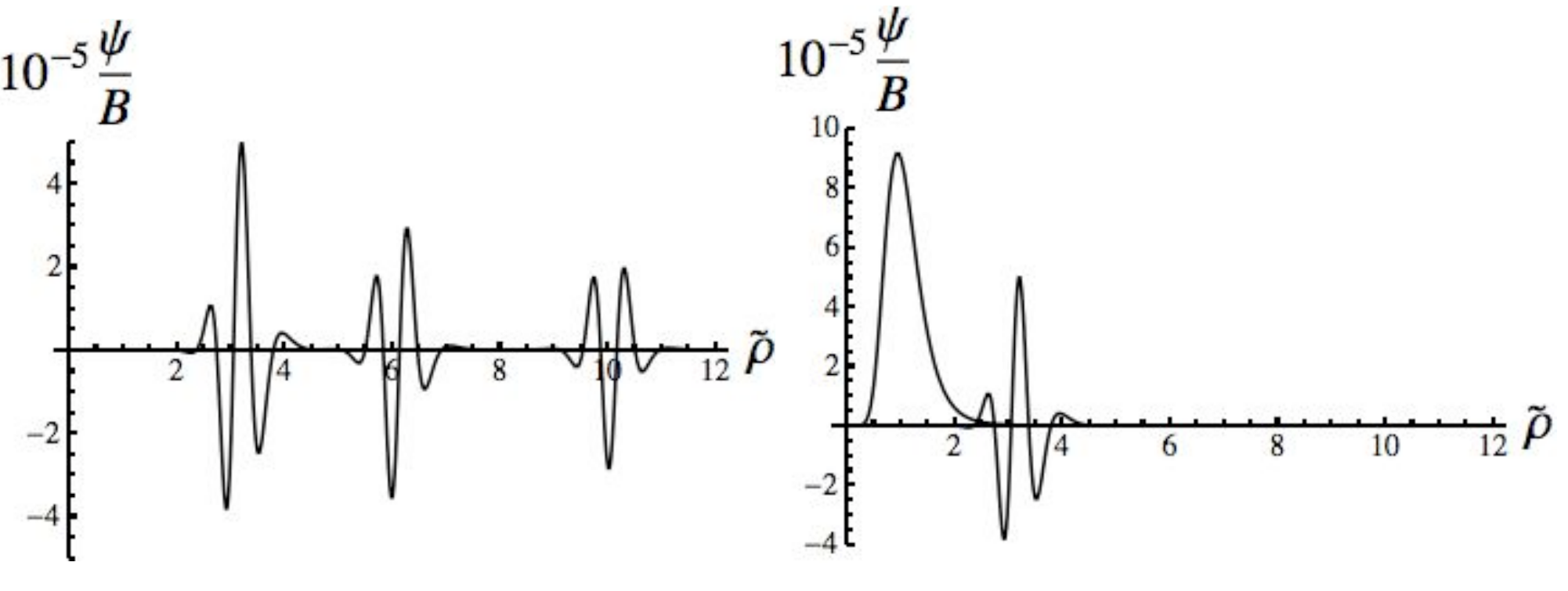} 
   \caption{\small  This shows $\ps/B$, the wave profiles for $m=10$ and $\vf=0$ at 5 different times at adapted scales. On the \lhs  $\ps/B$ is shown as it circulates inwards, from right to left at times $\tt=-10, -6$, and $-3$. The \rhs shows it as it circulates at closest approach to the axis $\tt=0$. Notice there is very little closer than $\tr\sim 0.4$ even at $\tt=0$. Superposed on the \rhs is the profile as the wave starts to retreat at $\tt=3$. Thereafter the profiles are the same as on the \lhs of the figure but in reverse order. At all times the wave profile centers around $\tr=\sq{1+\tt^2}$.}
   \end{figure}
This is greatest at $\tt=0$ exactly. Notice that there is {\it no time lag} between the wave arriving closest to the axis - most of the energy never gets nearer than $\tr\sim0.4$ (see Figure 1 at $\tt=0$) - and the resulting \ro of inertial frames. Just as in the work of Lindblom and Brill \cite{LB} on inertia induction from falling rotating spheres, the non-local effect appears instantaneously when   suitable Machian \coo (gauge) are chosen. This is a consequence of Mach's principle being embodied in the ``instantaneous" constraint \eeee of General Relativity. In electricity, the presence of a charge within a sphere at time $t$ can be ascertained ``instantaneously" by measuring the flux of its field out of a surrounding surface. You do not have to wait for the field to come out from the charge - it is there already. 

Returning to our \iii on the axis, a rod lying peacefully at the origin, initially oriented at $\vf=\vf_0$, will remain in the flat space near there and feel no torque, but, due to  the \ro of the \iii it will point towards $\vf=\phi(t)$ where 
\be
\phi(t)=\phi(0)+\int_{- \inf}^t <\!\! \om\!\!>_0dt=\phi(0)+ B^2 \fr{(2m+1)!}{2^{2m}}\LLL \arctan \tt + \fr{\pi}{2}+ \fr{\tt}{(2m+1)(1+\tt^2)}\RRR.
\lb{32}
\ee
When the whole pulse has passed it ends up oriented towards
\be
\phi(\inf)= \phi(0)+ B^2 \fr{(2m+1)!}{2^{2m}}\cd\pi.
\lb{33}
\ee
For symmetry we choose 
\be
\phi(0)= - B^2 \fr{(2m+1)!}{2^{2m}}\cd\fr{\pi}{2},
\lb{34}
\ee
so that $\phi=0$ when $\tt=0$ and
\be
\phi(t)= B^2 \fr{(2m+1)!}{2^{2m}}\LLL \arctan \tt  + \fr{\tt}{(2m+1)(1+\tt^2)}\RRR.
\lb{35}
\ee
\begin{figure}[h]
   \centering
      \includegraphics[width=12 cm]{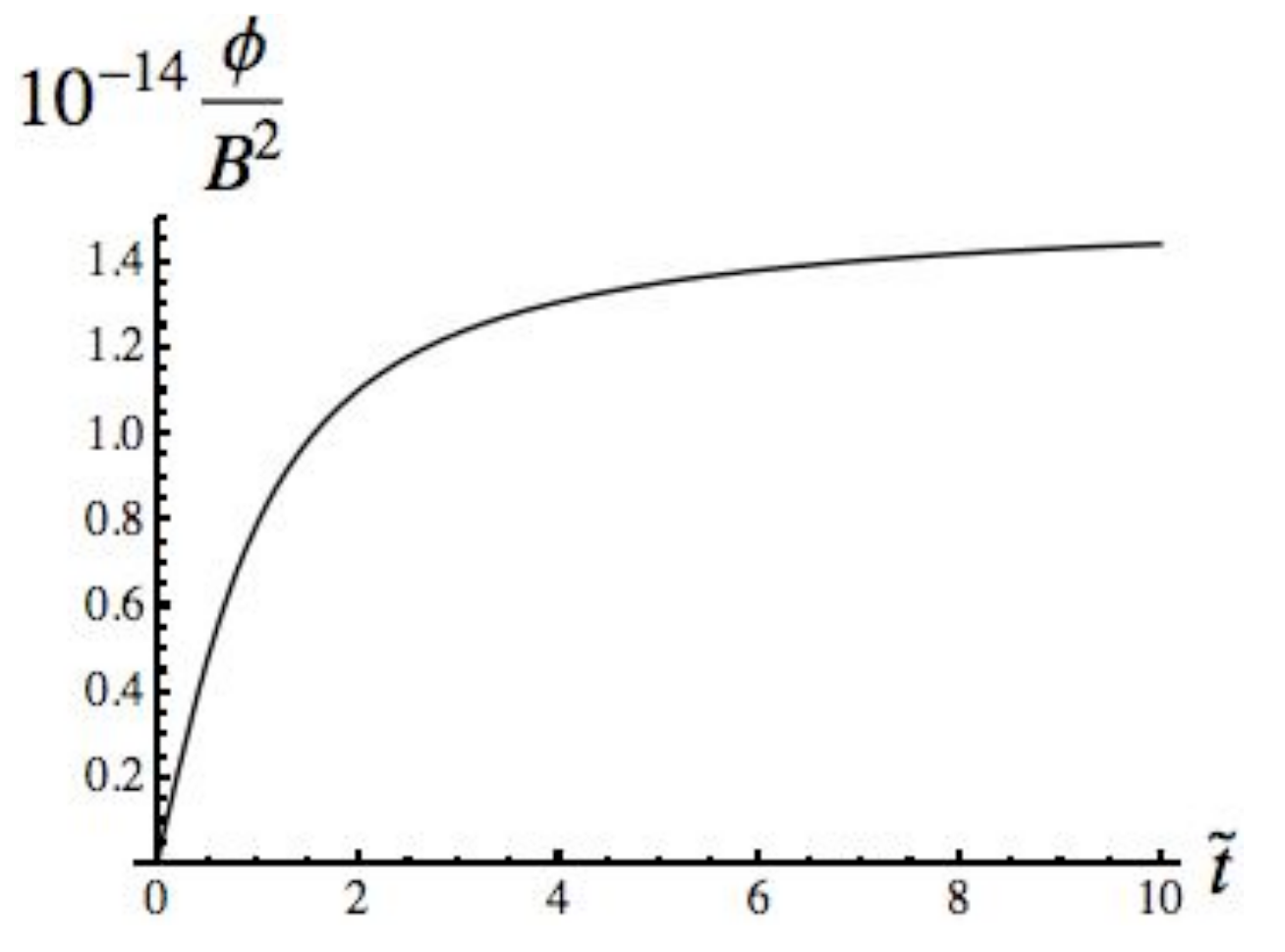} 
   \caption{\small  The angle $\phi(\tt)/B^2$ through which the \iii at the origin turns as a function of time for $m=10$. The initial angle before the wave comes in, is   chosen as $-\phi_0$ so that $\phi$ is zero at $\tt=0$. This produces  anti-symmetry about $\tt=0$. $\om(\tt,0)$  is given by the gradient of the curve. It is greatest at $\tt=0$.  }
   \end{figure}
Figure 2 shows $\phi(t)$ for $t\ge0$. It is antisymmeric about $t=0$.

 Far from the axis we have from (\ref{218}) and (\ref{A8})
\be
<\!\! \om\!\!>\sim\fr{B^2}{a}\cd\fr{m(m!)(2m-1)!!}{2^{m-1}}\cd\fr{1}{\tr^2}~~~~~~~,~~~~~~~\tr\gg1.
\lb{36}
\ee
As expected this is constant in time for   $\tr$ large, but falls off like $\tr^{-2}$  as befits a \cy system with a total \am per unit height, cf. \cite{BKL07},   
\be
J=  \fr{\pi}{\ka}aB^2\cd\fr{m(m!)(2m-1)!!}{2^{m-2}}. 
\lb{37}
\ee
  
For general $\tt$ and $\tr$, $<\!\om\!>$ is given in  (\ref{218}) with $I_m$ and $H_m$
from (\ref{A6}) and (\ref{A11}). But for $\tr^2\ll 1+\tt^2$ we need to show that our space is almost flat so that $<\!\om\!>$ can be properly interpreted as a \ro of the inertial frame. We therefore evaluate $<\!\om\!>$ not just at $\tr=0$, but in this larger region, and show that at each time $<\!\om\!>$ varies weakly with $\tr$ there. Indeed from (\ref{A16}) and (\ref{A17}) 
\be
<\!\!\om\!\!>\sim<\!\! \om\!\!>_0\LLL  1 - \fr{(2m-1)!!}{(m+1)!}\cd\fr{1}{(1+\tt^2)^{m+1}}\cd\fr{(2m-5)+(2m+7)\tt^2}{(m+1) +m\tt^2}\cd\lll\fr{\tr^2}{1+\tt^2}  \rrr^m  \RRR.
\lb{38}
\ee
 Figures 3 and 4 show $j$, $\ka$ times the \am density,  and the resulting $<\! \om\!(t,\rh)\!>$.
\begin{figure}[h]
   \centering
      \includegraphics[width=10cm]{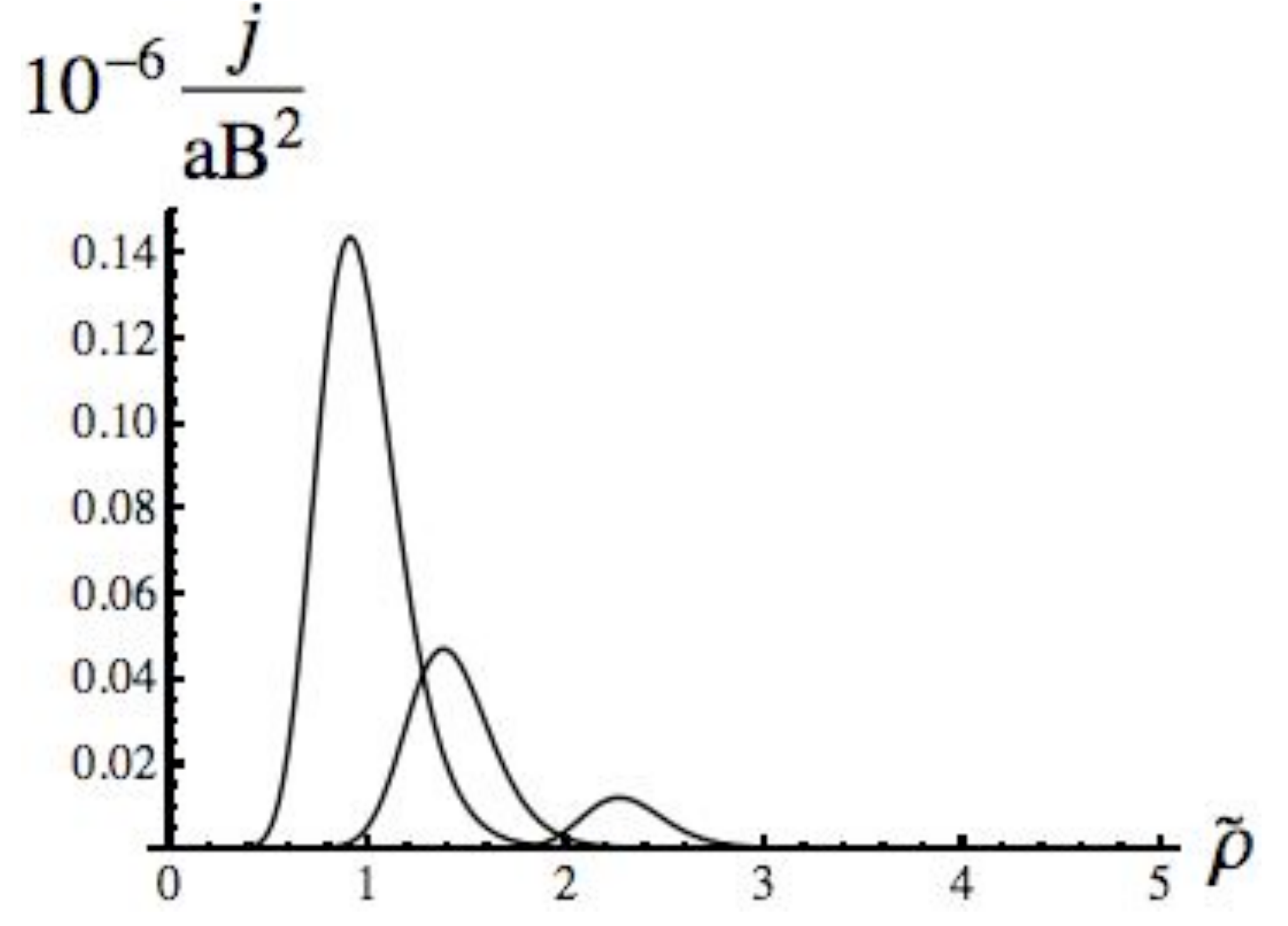} 
   \caption{\small  The profile of the \am density at $t\sim 0, 1$ and $2$. Again the drawing is for $m=10$.}
   \end{figure} 
   \begin{figure}[h]
   \centering
      \includegraphics[width=12cm]{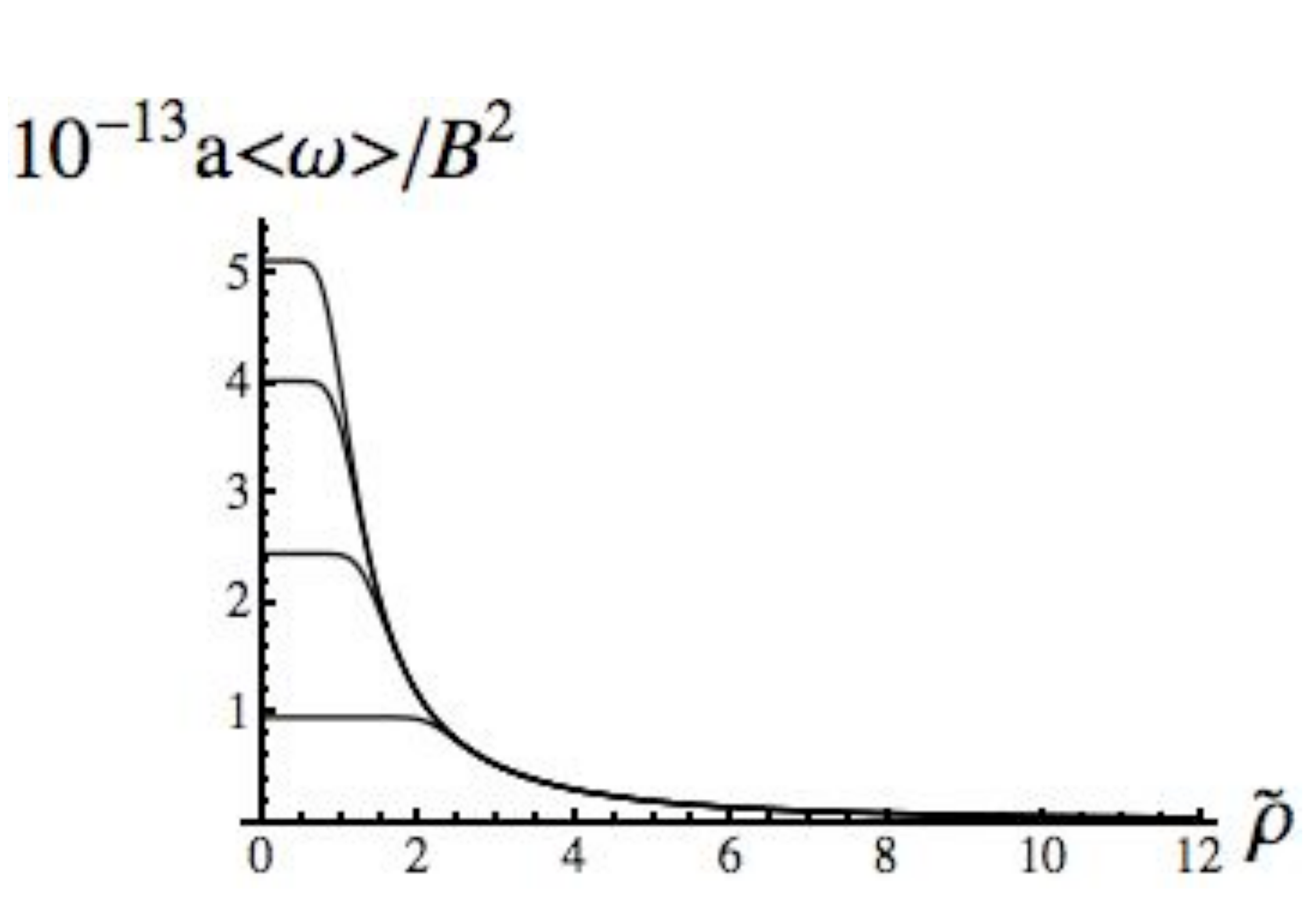} 
   \caption{\small  The profile $<\!\om\!>$  as a function of $\tr$ at selected times $\tt=0, \, 0.5, 1$, and $2$. Notice the still flat uniformly rotating central region when $\tt=0$.}
   \end{figure}
 The fact that the source terms in (\ref{23}) all vary as $\tr^{2m}$ shows that the other terms in the metric will vary similarly, so wherever $[\tr^2/(1+\tt^2)]^m$ is negligible our second order metric 
 will take the form
 \be
ds^2=e^{2\ps}\LLL   dt^2-d\rh^2-\rh^2\lll  d\vf \,- \!<\!\om\!>_0dt\rrr^2  \RRR -e^{-2\ps}dz^2~~~~~~,~~~~~~       \tr^2\ll(1+\tt^2)
\lb{39}
 \ee
with $\ps$ given by (\ref{210}). If we now go to rotating axes by writing $\td\vf=               \vf-\phi(t)$, then our metric becomes
\be
ds^2=e^{2\ps}\lll   dt^2-d\rh^2-\rh^2d\td\vf^2  \rrr -e^{-2\ps}dz^2~~~~~~,~~~~~~       \tr^2\ll(1+\tt^2)
\lb{310}
 \ee
which is flat in the reduced space. Furthermore the wave itself never penetrates significantly within $\tr/\sq{1+\tt^2}\sim 0.4$.
 
\sect{Angular momentum transport by gravitational torques}

Those who have studied \am transport in spiral galaxies, know that the classical gravitational  stress tensor is \cite{LK}, \cite{MB}
\be
\si_{kl}=\1k(2\di_k\ps\di_l \ps-\de_{kl}\sum_m|\di_m\ps|^2),  
\lb{41}
\ee
where here $\ps$ is the classical gravitational potential. The gravitational couple transfering \am outwards through a cylinder of radius $\rh$ is
\be
C_g=\int\ep_{3kl}X^k(\si^{lm}dS_m),
\lb{42}
\ee
where $d\vec S$ is the outward pointing surface element and $\vec X$ the radius vector   from the axis. Evidently this reduces to
\be
C_g=\fr{2}{\ka}\int\di_\rh\ps\di_\vf\ps \rh d\vf dz.
\lb{43}
\ee
For these gravitational stresses to carry \am outwards there must be a positive correlation between $\di_\rh\ps$ and $\di_\vf\ps$ averaged over such a cylinder. Such a correlation gives a trailing sense of spirality to contours of constant $\ps$ in the sense that the outer parts of a spiral galaxy trail that of the inner parts in the sense defined by the rotation. It is no different here; for \am transport outward, the spiral formed by contours of $\ps$ must trail, but contrarywise when they form a leading spiral with the outside further advanced than the inside, the \am is transported inward. Thus our knowledge of spiral structure in galaxies enabled us to predict that there would be leading spirals when the wave was going in and therefore transporting its \am inward, but trailing spirals when our \gw was receding from the axis. At $t=0$ there should be no \am transport so the contours of $\ps$ should form a \rog cartwheel structure with no spirality.  These expectations were beautifully confirmed when we plotted the contours of $\ps$  using Mathematica.
 \begin{figure}[h]
   \centering
 \includegraphics[width=14 cm]{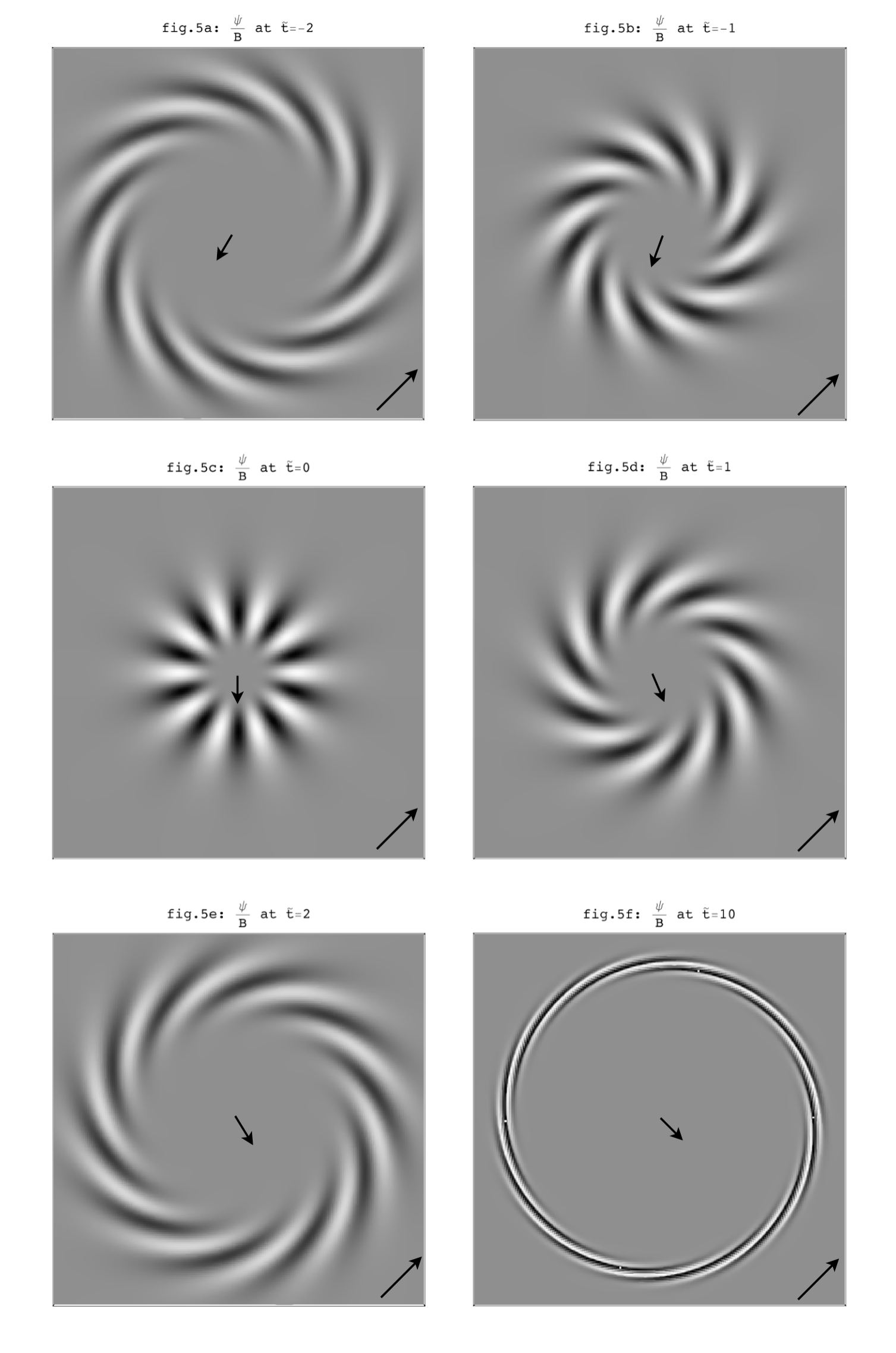} 
   \caption{\small This shows the $m=10$ wave which always   rotates anticlockwise. As it comes inwards (Figure 5a) at $\tt=-2$ it is in the form of a leading spiral with the outer parts of the arms ahead of the central parts. By $\tt=-1$  (Figure  5b) the spiral has started to open. By $\tt=0$  (fig.5c) the central parts have caught up and the spiral has changed to a cartwheel structure but rotation keeps it beyond  $\tr\sim0.4$.  By $\tt=1$  (Figure 5d) the spiral has become trailing as befits a wave that now feeds \am outwards. By $\tt=2$  (Figure 5e) the spiral becomes tighter    and the flat central cylinder becomes larger. We show $\tt=10$  (Figure 5f) at a small scale but notice the beautiful tight wrapping of the narrow arms. Also notice the opposite spirality of the conjugate pairs $\tt=\pm2$ and $\tt=\pm1$. The figures encompass a radius $\tr\sim 7$ ($\tr\sim   17$ for $\tt=10$). The height of $\ps/B$ reduced by a factor of $10^{-4}$ is between $0$ and $1$. Lighting falls at $45^0$ from the left. The view is along the $z$ axis from above at a distance of $10^{-4}\ps/B=40$.}
   \end{figure}
 The waves always rotate counterclockwise in Figure 5. 
 They start as quite tightly wrapped leading spirals; as they move inwards they keep rotating but the wrapping becomes less tight. At $t=0$ they unwrap to form a \rog cartwheel structure and thereafter they move outward as trailing spirals carrying \am with them. It is when they are closest in that their \ro causes the greatest \ro of the \iii in the central region,  though the waves themselves never reach much further in than $\tr=1$. Nevertheless the second order effect of the \am causes the \ro of the \iii within. This is illustrated by the direction of the central arrow that shows the orientation of a rod initially at rest at the origin. Because this effect is second order, the \ro  seen in the figure is dependent on the amplitude of the wave, whereas the spiral shapes are not.
 \sect{Energy and \am}
In Section 3 we used the asymptotic form of $<\! \om\!>$
 to deduce the total \am per unit height but we could also calculate it via the \am density $j/\ka$ 
 \be
 J=\1k\int_0^\inf j 2\pi \rh d\rh=\fr{4\pi}{\ka}aB^2m[(2m-1)!!]^2\LLL mI_{m-1}+(2m+1)\tt^2I_m  \RRR= \fr{\pi}{\ka}aB^2\fr{m(m!)(2m-1)!!}{2^{m-3}}. 
 \lb{51}
 \ee
Any quantity $\ps$
 that obeys $\square \ps=0$ automatically conserves both $J$ and the quantity 
 \be
 \1k\int(\dot\ps^2+|\vna\ps|^2)  dV.
 \lb{52}
 \ee
The identity
\be
 \1k\di_t(\dot\ps^2+|\vna\ps|^2)=\1k\vna\cd(2\dot\ps\vna\ps)
 \lb{53}
\ee
follows directly from the wave \eee and can be interpreted as a conservation law with $\fr{2}{\ka}\dot\ps\vna\ps\cd d\vec S$ being the flux through an element of the closed surface surounding a volume $V$. We notice that the flux vector now directed through a surface $\vf=$const  will on multiplication by $\rh$ give the \am density $j/\ka$. 
Integrating  $\1k(\dot\ps^2+|\vna\ps|^2)$ over a volume of unit height we find the grand total  and using the formul\ae\, of the Appendix,
\be
\int_0^\inf\int_0^{2\pi}\1k(\dot\ps^2+|\vna\ps|^2)d\vf\rh d\rh=\fr{\pi}{\ka} B^2\fr{m!(2m+1)!!}{2^{m-1}}, 
\lb{54}
\ee
which is independent of time as expected. We may identify this conserved quantity with energy per unit length via the following argument. The conservation laws $\na_b\TT^b_a=0$, see \cite{BKL07}, in the reduced $2+1$ space follow from its contracted Bianchi identities of which the $x^0=t$ one reads
\be
\fr{1}{\rh}\LLL  (\rh \TT^0_0)^{.}+ (\rh \TT^1_0)'  + \di_\vf(\rh \TT^2_0)    \RRR=0.
\lb{55}
\ee
Ordinary derivatives in flat space have replaced covariant ones because $\TT^b_a$ is already of second order in the wave amplitude:
\be
\ka \TT_{ab}= \RR_{ab} - \ts{\fr{1}{2}} g_{ab}\RR=2\di_a\ps\di_b\ps - g_{ab}\lll \dot\ps^2-|\vna\ps|^2  \rrr,
\lb{56}
\ee
where $\vna\ps$ is the normal flat space gradiant in $\rh$ and $\vf$,
\be
|\vna\ps|^2= {\ps'}^2+\fr{1}{\rh^2}(\di_\vf\ps)^2.
\lb{57}
\ee
From (\ref{56}) we see that
\be
\ka\TT^0_0=\dot\ps^2+|\vna\ps|^2
\lb{58}
\ee
which is the density of our conserved quantity. The total energy per unit length as measured from the conicity at infinity, $\ga_\inf$, was shown to be $2\pi\ga_\inf/\ka$ by \cite{ABS}. 

In \cite{BKL07}, eq. (5.31), we show that  $2\pi\ga_\inf/\ka$ is given by the \lhs of (\ref{54}). This completes the identification of our conserved quantity with the energy per unit height. It also shows that the wave  produces a long range effect  on $\ga$  far from the region $\rh\sim t$ where its main amplitude lies. It is one of the strange properties of \cy \GR that this conicity is independent of $\rh$ outside the wave. This has the effect that while particle trajectories are globally bent by the defect angle nevertheless there are no   orbits along finite sections of circles (about the axis) caused by the long distance gravity of the \cy wave. Rather the geodesics are like the geodesics on a cone all of which go to infinity. It would be interesting to calculate the gravity due to a more realistic three dimensional wave pulse for which the odd behaviour at infinity found in 2+1 dimensions would not obscure the interpretation \cite{LL}.

\vs
  \Large{\bf Acknowledgements}
 \vskip .5 cm
 \normalsize 
 \setlength{\baselineskip}{20pt plus2pt}
  
 J.B. and J.K. are grateful to the Institute of Astronomy, Cambridge University, and the Royal Society for hospitality and support.  J.B. also acknowledges the hospitality of the Albert Einstein Institute in Golm    and the Institute of Theoretical Physics at FSU in Jena, the support of the Alexander von Humbolt Foundation, the partial support from the Grant GA\v CR 202/06/0041 of the Czech Republic, of Grant No  LC06014 and MSM0021620860 of the Ministry of Education  and from SFB/TR7 in Jena.

\vskip .5 cm
\begin{appendix}
\setcounter{equation}{0}
\section{ Appendix: Evaluation of the Integrals $I_m$ and $H_m$  }
\renewcommand{\theequation}{\Alph{section}.\arabic{equation}}
\normalsize
\setlength{\baselineskip}{20pt plus2pt}
  \vs
\be
I_m=\int_0^u u^mQ^{-m-3/2}du~~~;~~~H_m=\int_u^\inf u^{m-1}Q^{-m-3/2}du.
\lb{A1}
\ee
We write 
\be
Q=u^2+2bu+c^2~~~,{\rm where}~~~b=1-\tt^2~~~{\rm and} ~~~c=1+\tt^2,
\lb{A2}
\ee
so $c\ge b$. Then
\be
\int_0^uQ^{-1/2}du={\rm arc}\sinh\lll \fr{u+b}{\sq{c^2-b^2}}  \rrr - {\rm arc}\sinh\lll \fr{b}{\sq{c^2-b^2}}  \rrr,
\lb{A3} 
\ee
and
\be
\int_0^uQ^{-3/2}du=-2\fr{\di}{\di c^2}\lll  \int_0^uQ^{-1/2}du   \rrr=\fr{u+b}{c^2-b^2}Q^{-1/2}-\fr{b/c}{c^2-b^2}\eq X(b,u).
\lb{A4}
\ee
Now differentiate $m$ times with respect to $b$:
\be
(-1)^m (2m+1)!!\int_0^u u^mQ^{-m-3/2}du=\fr{\di^m X(b,u)}{\di b^m}.
\lb{A5}
\ee
So
\be
I_m(u)=  \fr{(-1)^m}{(2m+1)!!}\fr{\di^m X(b,u)}{\di b^m},
\lb{A6} 
\ee
where in this expression $c=1+\tt^2$ and, after differentiation $b=1-\tt^2$. Now from (\ref{A4}) 
\be
X(b,\inf)=\fr{1-b/c}{c^2-b^2}=\fr{1}{c(c+b)}, 
\lb{A6a}
\ee 
so
\be
I_m(\inf)=\fr{m!}{(2m+1)!!}\cd\fr{1}{c(c+b)^{m+1}}.
\lb{A7}
\ee
For general $b$ and $u$  
\be
\int_u^\inf Q^{-3/2}du= \fr{1}{c^2-b^2}\LLL   1-(u+b)Q^{-1/2}      \RRR.
\lb{A8}
\ee 
Differentiating with respect to $c^2$ and multiplying by $-2/3$:
\be
\int_u^\inf Q^{-5/2}du= \fr{1}{3(c^2-b^2)^2}\LLL   2-\fr{(u+b)(2Q+c^2-b^2)}{Q^{3/2}} \RRR\eq Y(b,u).
\lb{A9}
\ee
Differentiating $m-1$ times \wrt $b$ we find
\be
(-1)^{m-1}\fr{1}{3}(2m+1)!!\int_u^\inf u^{m-1}Q^{-m-3/2}du=\fr{\di^{m-1}Y(b,u)}{\di b^{m-1}}.
\lb{A10}
\ee
Hence,
\be
H_m=  \fr{3(-1)^{m-1}}{(2m+1)!!}\fr{\di^{m-1}Y(b,u)}{\di b^{m-1}},
\lb{A11}
\ee
again $b$ must be set equal to $1-\tt^2$ after the differentiation. From (\ref{A9}) we see that 
\be
Y(b,0)=\fr{1}{3c^2}\LLL   \fr{1}{c(c+b)}+\fr{1}{(c+b)^2} \RRR.
\lb{A12}
\ee
So
\be
(-1)^{m-1}\fr{\di^{m-1} Y(b,0)}{\di b^{m-1}}=\fr{1}{3c^2}\LLL   \fr{(m-1)!}{c(c+b)^m}+\fr{m!}{(c+b)^{m+1}} \RRR
\lb{A13}
\ee
and
\be
H_m(0)= \fr{(m-1)!}{(2m+1)!!}\cd \fr{b+(m+1)c}{ c^3(c+b)^{m+1}}.
\lb{A14}
\ee
Whereas (\ref{A6}) and (\ref{A11}) give $I_m$ and $H_m$ for all values of $u$, 
(\ref{A7}) and (\ref{A14}) give their values at $0$ and $\inf$, nevertheless it is useful to view their behaviours at small and at large $u$ explicitly. In particular for $\tr^2\ll\fr{1}{2}(1+\tt^2)$, the region encompassed by the wave but not `feeling' it, we may expand $Q^{-m-3/2}$ in the form
\be
Q^{-m-3/2}\sim \fr{1}{c^{2m+3}}\LLL    1-(2m-3)\cd\fr{bu}{c^2}       \RRR~~~~~~~,~~~~~~~~~~~~~~~~u\ll c~.
\lb{A15}
\ee
Then 
\be
I_m(u)\sim \fr{u^{m+1}}{c^{2m+3}}\LLL    \fr{1}{m+1}-  \fr{ 2m-3}{m+2}\cd\fr{bu}{c^2}       \RRR~~~~~~~,~~~~~~~~~~~~u\ll c~,
\lb{A16}
\ee
and likewise
\be
H_m(u)\sim H_m(0)-  \fr{u^m}{c^{2m+3}}\LLL      \fr{1}{m}-  \fr{ 2m-3}{m+1}\cd\fr{bu}{c^2}            \RRR~~~~~~~,~~~~~~~u\ll c~.
\lb{A17}
\ee
Likewise for $\tr^2\gg\fr{1}{2}(1+\tt^2)$ which is $u\gg c$ 
\be
Q^{-m-3/2}\sim \fr{1}{u^{2m+3}}\LLL    1-(2m-3)\fr{b}{u}       \RRR~~~~~~~~~~~~~~~~~~~~~~~~~,~~~~~~~u\gg c~,
\lb{A18}
\ee
so
\be
I_m(u)\sim I_m(\inf) - \fr{1}{u^{m+2}}\LLL \fr{1}{m+2}-  \fr{ 2m-3 }{m+3}\cd\fr{b}{u}           \RRR~~~~~~~~~~~,~~~~~~~~~u\gg c~,
\lb{A19}
\ee
and 
\be
H_m(u)\sim \fr{1}{u^{m+3}}\LLL    \fr{1}{m+3}-\fr{2m-3}{m+4}\cd\fr{b}{u}\RRR~~~~~~~~~~~~~~~,~~~~~~~~~~~u\gg c~.
\lb{A20}
\ee
Finally when $b=c$, $Q$ becomes a perfect square so the integrals are easier; since both $b$  and $c$ are then $1$ we express the answer that way:
\be
I_m=\sum_{s=0}^m (-1)^{m-s} {m\choose s}   \fr{(1+u)^{s-2m-2}-1}{(s-2m-2)}  ~~~~,~~~~H_m=\sum_{s=0}^{m-1}  (-1)^{m-s} {m\choose s} \fr{ (1+u)^{s-2m-2}}{(s-2m-2)}.
\lb{A21}
\ee

\end{appendix}

  \end{document}